\RequirePackage{fix-cm}
\documentclass[smallextended]{svjour3}       
\smartqed  
\usepackage{graphicx}
 \journalname{Foundations of Science}
\usepackage{amssymb}
\begin{document}

\title{Adversus singularitates: The ontology of space-time singularities
}

\titlerunning{Ontology of space-time singularities}        

\author{Gustavo E. Romero}


\institute{Instituto Argentino de Radioastronom{\'{i}}a (IAR, CCT La Plata, CONICET) \at
              C.C. No. 5, 1894, Villa Elisa, Buenos Aires, Argentina. \\
              Tel.: +54-221-482-4903\\
              Fax: +54-221-425-4909\\
              \email{romero@iar-conicet.gov.ar}
}

\date{Received: date / Accepted: date}

\maketitle

\begin{abstract}

I argue that there are no physical singularities in space-time. Singular space-time models do not belong to the ontology of the world, because of a simple reason: they are concepts, defective solutions of Einstein's field equations. I discuss the actual implication of the so-called singularity theorems. In remarking the confusion and fog that emerge from the {\em reification} of singularities I hope to contribute to a better understanding of the possibilities and limits of the theory of General Relativity. 

\keywords{Ontology \and space-time \and general relativity \and singularities}
 \PACS{ 01.70.+w  \and 04.20.Gz }
\end{abstract}

\begin{quotation}

Einstein always was of the opinion that singularities in classical field theory are intolerable. They are intolerable from the point of view of classical field theory because a singular region represents a breakdown of the postulated laws of nature. I think one can turn this argument around and say that a theory that involves singularities, and involves them unavoidably, carries within itself the seeds of its own destruction...\\
\begin{flushright}
{\rm Peter G. Bergmann}
\end{flushright}
\end{quotation}

\section{Introduction}

The attitude of physicists towards singular solutions of classical field theories, in particular theories of gravitation, has changed along time. In the case of General Relativity (GR) this change went from plain rejection\footnote{``For a singularity brings so much arbitrariness into the theory that it actually nullifies its laws...Every field theory, in our opinion, must therefore adhere to the fundamental principle that singularities of the field are to be excluded'', Einstein and Rosen (1935).} to {\em reification}\footnote{{\em Reification} is the mental process that elevates a concept to the status of an existent entity. See, for instance, this statement of Hawking and Ellis (1973): ``Recent observations of the microwave background indicate that the universe contains enough matter to cause a time-reversed closed trapped surface. This implies the existence of a singularity in the past, at the beginning of the present epoch of expansion of the universe. This singularity is in principle visible to us.''}. After significant advances in the understanding of singular space-time models during the 1990s, made by both physicists and philosophers (Clarke 1993, Earman 1995), an attitude of caution, even skepticism, has emerged.  

This paper is a plea for intolerance about singularities. I shall argue that Einstein's attitude towards singular solutions of GR was well-motivated and essentially correct. My thesis can be simply stated: {\em there are no singularities}. There are only singular space-time models, and they are defective representations of reality. 

My {\it blitz} against singularities will follow a simple plan. I start in the next section briefly discussing the relation between conceptual ways of representation and reality. Then, I formulate a simple ontological basis of GR. In the subsequent sections I review the standard definitions of singular space-times, and I discuss the ontological import of such models. If there are no singularities in GR, what are its main referents? A final section is devoted to this issue. I close reviewing the highlights and prospects.

\section{Language, representation, and reality}\label{representation}

We all make a basic assumption when doing science: there is a reality to be known. Without the postulate of the independent existence of a real world the scientific enterprise would be vain. I shall not discuss this primary assumption here. Rather, I want to focus on how we represent reality in our attempts to understand it. 

In order to build representations of the world we use conceptual systems called {\em languages}. In ordinary life natural languages such as English, German or Spanish are, or seem to be, enough. If we want to penetrate deeper into the structure of reality, however, we need formal and less vague languages like those provided by logic and mathematics. 

Essentially, a formal language is a system of signs with a set of explicit rules to generate valid combinations of symbols (see, for instance, the treatises by Bunge 1974a,b and Martin 1978). These rules give instructions about how valid arrangements of symbols (called {\em formulas}) are formed ({\em syntactic rules}), or they relate symbols and formulas with extra-linguistic objects ({\em semantic rules}). The operation of {\em deduction} allows to obtain valid formulas from valid formulas. If a set of formulas is closed under deduction, we call it a {\em theory}. Any {\em interpreted}\footnote{`Interpreted' in this context means `endowed with semantic rules'.} theory, with the help of auxiliary conditions, should produce {\em statements} about states of affairs that occur in the world. If a state of affairs can be used to validate a statement and this statement satisfies truth conditions\footnote{See Tarski (1983) formal truth.} in the theory, we say that the theory (actually one of its models) {\em represents} some aspect of the world. 

The use of formal languages in science brings many advantages, to the point that those specific scientific disciplines that do not make intensive use of formal methods are generally underdeveloped. Clarity and precision are gained through formalization. This results in a significant reduction of the vagueness that is inherent to natural languages. The extensive use of formalized languages, besides, enables us to elicit in a systematic way the consequences of our assumptions. The adoption of the special class of formalized languages of mathematics makes possible to introduce quantitative and complex representations of the properties and changes we detect in the world. 

A basic assumption of factual science is that a property can be represented by some mathematical function. Reality is not mathematical, but certainly our more accurate representations of it are mathematical. Physical systems, in general, are described by models where properties and processes are represented by mathematical constructs. Models, in this way, are representations of the mechanisms we assume occur in physical systems. This is so to the extent that {\em to explain} a thing is to unveil the mechanisms that operate in it, i.e. to faithfully represent the manifold of physical processes with a coherent system of mathematical functions and constructs (Bunge 2006)

Sometimes, in highly elaborated theories, however, formalization can reach such a degree of complexity that the semantics of the language might be difficult to elucidate. This results in problems of interpretation and is a source of confusion. Quantum mechanics and quantum field theories, for instance, are highly formal theories whose interpretations are still matter of strong controversy. There is no agreement even on what the referents of these theories are. This is an unfortunate situation that should be settled by logic, semantical analysis, and even experiment (see, e.g., Bunge 1967; Perez-Bergliaffa et al. 1993, 1996).

General Relativity is not free of this sort of problems. The complexity of the field equations, which are non-linear, and the interpretation of the metric tensor field have resulted in concerns on the ontological assumptions of the theory. General Relativity is one of the several, and certainly the most important, theories of the geometrization program of physics launched in the second decade of the 20th century (for a review of the program see Vizgin 1994). Although several first-rate theoreticians, such as  Weyl, considered GR as a theory of space-time, susceptible, in principle, to be expanded to encompass also electromagnetic phenomena, Einstein initial attitude from 1915 and up to 1919 was more cautious. In a letter to Sommerfeld written towards the end of 1915, Einstein said (cited in Vizgin 1994):
\begin{quotation}
[GR] essentially gives a law of the gravitational field, doing this, moreover, completely uniquely if the requirement of general covariance is satisfied.
\end{quotation}    
Several others comments by Einstein from those years provide clear evidence that he considered GR as a theory of the gravitational field, that should be modified somehow to include quantum effects. He even noticed that (Einstein 1916):
\begin{quotation}
An atom must, because of the intra-atomic motion of the electron, emit not only electromagnetic but also gravitational energy, albeit in a negligible amount. Since nothing like this occurs in reality, it is evident that quantum theory must modify not only Maxwellian electrodynamics but also the new theory of gravitation [GR].
\end{quotation}

Although at the end of the decade of 1910 Einstein thought of the metric of space-time as a tensor field whose role in GR was to represent the gravitational potential, as time went by he shifted towards a full geometrization of physics. He started then to look for suitable generalizations of GR that could accommodate electrodynamics. He also searched for particle-like solutions of different field equations in order to accomplish the incorporation of the discreteness required by the quantum aspects of reality (van Dongen 2010). His intolerance to space-time singularities remained, nonetheless, unshakable. 

\section{What is space-time?}\label{space-time}    

General relativity is said sometimes to be a `a theory of space and time'. We read, for instance, in the classic textbook by Misner, Thorne, and Wheeler (1973): ``Space acts on matter, telling it how to move. In turn, matter reacts back on space, telling it how to curve"'. The concept of space-time, however, is presupposed by all classical field theories\footnote{And most quantum theories. One exception is quantum loop gravity, which, up to some extent, can be considered as an ontological theory. See, e.g., Rovelli (2004).}. The concept was introduced by Minkowski (1908), and belongs more to ontology than to physics (Romero 2012a). A formal construction of space-time can be obtained starting from an ontological basis of either things (Perez-Bergliaffa et al. 1998) or events (Romero 2012b). In what follows I provide a simple outline of the relation between space-time and GR.

The basic ontological assumption is:\\

P0 - Ontological. {\em Space-time is the ontological composition of all events}. \\

Events can be considered as primitives or can be derived from things as changes in their properties if things are taken as ontologically prior. Both representations are equivalent since things can be construed as bundles of events (Romero 2012b). Since composition is not a formal operation but an ontological one\footnote{For instance, a human body is composed by cells, but is not just a mere collection of cells since it has emergent properties and specific functions far more complex than those of the individual components.}, space-time is not a concept nor an abstraction, but an emergent entity. As any entity, space-time can be represented by a concept. The usual representation of space-time is given by a 4-dimensional manifold $E$ equipped with a metric field $g_{ab}$:

$$  
{\rm ST}\hat{=}\left\langle E, g_{ab}\right\rangle.
$$

I insist: space-time {is not} a manifold (i.e. a mathematical construct) but the ``totality'' of all events. A specific model of space-time requires the specification of the source of the metric field. This is done through another field, called the ``energy-momentum'' tensor field $T_{ab}$. Hence, a model of space-time can be denoted by the following triplet:

$$  
M_{\rm ST}=\left\langle E, g_{ab}, T_{ab}\right\rangle.
$$

The relation between both tensor fields is given by Einstein's field equations. The metric field represents the gravitational potential. The energy-momentum field, the potential of change in space-time. 

We can summarize all this in the following axioms.\\

${\rm P1 - Syntactic}.$ The set $E$ is a $C^{\infty}$ differentiable, 4-dimensional, real pseudo-Riemannian manifold.\\

${\rm P2 - Syntactic}. $ The metric structure of $E$ is given by a tensor field of rank 2, $g_{ab}$, in such a way that the differential distance $ds$ between two events is: $$ds^{2}=g_{ab} dx^{a} dx^{b}.$$

${\rm P3 - Syntactic}.$ The tangent space of $E$ at any point is Minkowskian, i.e. its metric is given by a symmetric tensor $\eta_{ab}$ of rank 2 and trace $-2$.\\

${\rm P4 - Syntactic}.$ The metric of $E$ is determined by a rank 2 tensor field $T_{ab}$ through Einstein's field equations:

\begin{equation}
G_{ab}-g_{ab}\Lambda=\kappa T_{ab}, \label{Eq-Einstein} 
\end{equation}
where $G_{ab}$ is the Einstein's tensor (a function of the second derivatives of the metric) and $\Lambda$ is a constant.\\

${\rm P5 - Semantic}.$ The elements of $E$ represent physical events.\\

${\rm P6 - Semantic}.$ Space-time is represented by and an ordered pair $\left\langle E, \; g_{ab}\right\rangle$: $${\rm ST}\hat{=}\left\langle E, g_{ab}\right\rangle.$$

${\rm P7 - Semantic}.$ There is a non-geometrical field represented by a 2-rank tensor field $T_{ab}$ on the manifold E.\\ 

${\rm P8 - Semantic}.$ A specific model of space-time is given by: $$M_{{\rm ST}}=\left\langle E, g_{ab}, T_{ab}\right\rangle.$$

Up to here we have a purely ontological theory. To transform this into a physical theory (GR) we need three additional semantical postulates of physical nature. \\

${\rm P9 - Semantic}.$ The metric field $g_{ab}$ represents the potential of the gravitational field.  \\ 

${\rm P10 - Semantic}.$ The energy-momentum tensor field $T_{ab}$ represents the energy-momentum distribution of the physical systems other than the gravitational field.\\ 

${\rm P11 - Semantic}.$ The constant $\kappa$ gives the strength of the coupling between the gravitational and non-gravitational fields. Its value is $\kappa=8\pi G c^{-4}$, where $G$ and $c$ are the gravitational and vacuum speed of light constant, respectively. \\

Some comments are in order. First, with the interpretation given by the last 3 postulates, Eqs. (\ref{Eq-Einstein}) express a law of nature: the law of gravitational field. The equations, just a formal constraint so far, become Einstein's field equations. Second, the energy-momentum tensor expresses the potential of a physical system to change. The more a system can change, the more energy it has. Energy, then, is not a thing, but a property: the most general of all properties, shared by all things. It is the capability of changing (Bunge 1977). Changes are events, so energy reflects the structure of space-time (the composition of all events). Finally, singularities are features of some solutions of Eqs. (\ref{Eq-Einstein}). Hence, they appear in our representations of ST and not in space-time itself.

\section{Singular space-time models}\label{singularities-I}

A space-time model is said to be {\sl singular} if the manifold $E$ defined in the previous section is {\sl incomplete}. A manifold is incomplete if it contains at least one {\sl inextendible} curve. A curve $\gamma:[0,a)\longrightarrow E$ is inextendible if there is no point $p$ in $E$ such that $\gamma(s)\longrightarrow p$ as $a\longrightarrow s$, i.e. $\gamma$ has no endpoint in $E$. A given space-time model $\left\langle E, \;g_{ab}\right\rangle$ has an {\sl extension} if there is an isometric embedding $\theta: M\longrightarrow E^{\prime}$, where $\left\langle E^{\prime}, g_{ab}^{\prime}\right\rangle$ is another space-time model and $\theta$ is an application onto a proper subset of $E^{\prime}$. A {\em singular} space-time model contains a curve $\gamma$ that is inextendible in the sense given above. Singular space-times are said to contain singularities, but this is an abuse of language: singularities are not `things' in space-time, but a pathological feature of some solutions of the fundamental equations of the theory.  


Several singularity theorems can be proved from pure geometrical properties of the space-time model (Clarke 1993). The most important of these theorems is due to Hawking and Penrose (1970):\\

{\bf Theorem.} Let $\left\langle E,\;g_{ab}\right\rangle$ be a time-oriented space-time satisfying the following conditions:
\begin{enumerate}
	\item $R_{ab}V^{a}V^{b}\geq 0$ for any non space-like $V^{a}$\footnote{$R_{ab}$ is the Ricci tensor obtained by contraction of the curvature tensor of the manifold $E$.}.
	\item Time-like and null generic conditions are fulfilled.
	\item There are no closed time-like curves.
	\item At least one of the following conditions holds
	
\begin{itemize}
	\item a.  There exists a compact\footnote{A space is said to be compact if whenever one takes an infinite number of "steps" in the space, eventually one must get arbitrarily close to some other point of the space. Thus, whereas disks and spheres are compact, infinite lines and planes are not, nor is a disk or a sphere with a missing point. In the case of an infinite line or plane, one can set off making equal steps in any direction without approaching any point, so that neither space is compact. In the case of a disk or sphere with a missing point, one can move toward the missing point without approaching any point within the space. More formally,  a topological space is compact if, whenever a collection of open sets covers the space, some sub-collection consisting only of finitely many open sets also covers the space. A topological space is called compact if each of its open covers has a finite sub-cover. Otherwise it is called non-compact. Compactness, when defined in this manner, often allows one to take information that is known locally -- in a neighborhood of each point of the space -- and to extend it to information that holds globally throughout the space.} achronal set\footnote{A set of points in a space-time with no two points of the set having time-like separation. } without edge.
	\item b. There exists a trapped surface.
	\item c. There is a $p\in E$ such that the expansion of the future (or past) directed null geodesics through $p$ becomes negative along each of the geodesics.  
\end{itemize}
\end{enumerate}
 
Then, $\left\langle E,\;g_{ab}\right\rangle$ contains at least one incomplete time-like or null geodesic. \\

If the theorem has to be applied to the physical world, the hypothesis must be supported by empirical evidence. Condition 1 will be satisfied if the energy-momentum $T^{ab}$ satisfies the so-called {\em strong energy condition}: $T_{ab}V^{a}V^{b}\geq -(1/2)T^{a}_{a}$, for any time-like vector $V^{a}$. If the energy-momentum is diagonal, the strong energy condition can be written as $\rho+3 p\geq 0$ and $\rho + p\geq 0$, with $\rho$ the energy density and $p$ the pressure. Condition 2 requires that any time-like or null geodesic experiences a tidal force at some point in its history. Condition 4a requires that, at least at one time, the universe is closed and the compact slice that corresponds to such a time is not intersected more than once by a future directed time-like curve. The trapped surfaces mentioned in 4b refer to horizons due to gravitational collapse.  Condition 4c requires that the universe is collapsing in the past or the future. 
 
The theorem is purely geometric, no physical law is invoked. Theorems of this type are a consequence of the gravitational focusing of congruences. An outline of the proof of the theorem is given in the Appendix.

\section{Are there singularities?}\label{singularities-II}

Singularity theorems are not theorems that imply physical existence, under some conditions, of space-time singularities. Material existence cannot be formally implied. Existence theorems imply that under certain assumptions there are functions that satisfy a given equation, or that some concepts can be formed in accordance with some explicit syntactic rules. Theorems of this kind state the possibilities and limits of some formal system or language. The conclusion of the theorems, although not obvious in many occasions, are always a necessary consequence of the assumptions made. 

In the case of singularity theorems of classical field theories like GR, what is implied is that under some assumptions the solutions of the equations of the theory are defective beyond repair. The correct interpretation of these theorems is that they point out the {\em incompleteness} of the theory: there are statements that cannot be made within the theory. In this sense (and only in this sense), the theorems are like G\"odel's famous theorems of mathematical logic\footnote{G\"odel's incompleteness theorems are two theorems of mathematical logic that establish inherent limitations of all but the most trivial axiomatic systems capable of doing arithmetic. The first theorem states that any effectively generated theory capable of expressing elementary arithmetic cannot be both consistent and complete (G\"odel 1931). The second incompleteness 
theorem, shows that within such
a system, it cannot be 
demonstrated its own 
consistency.}.  

To interpret the singularity theorems as theorems about the existence of certain space-time models is wrong. Using elementary second order logic is trivial to show that there cannot be non-predicable objects (singularities) in the theory. If there were a non-predicable object in the theory,
\begin{equation}
	\left(\exists x\right)_{E} \; \left(\forall P\right) \sim Px, \label{P}
\end{equation}
where the quantification over properties in unrestricted. The existential quantification $\left(\exists x\right)_{E}$, on the other hand, means

$$\left(\exists x\right)_{E} \equiv \left(\exists x\right) \wedge \left( x\in E\right).$$ 

Let us call $P_{1}$ the property `$x\in E$'. Then, formula (\ref{P}) reads:
\begin{equation}
\left(\exists x\right)	\left(\forall P\right) (\sim Px \; \wedge P_{1}x ), \label{P1}
\end{equation}
which is a contradiction, i.e. it is false for any value of $x$. 

We conclude that there are no singularities nor singular space-times. There is just a theory with a restricted range of applicability.

\section{The ontology of General Relativity}\label{ontology}

The reification of singularities can lead to accept an incredible ontology. We read, for instance, in a book on foundations of General Relativity cite{Kriele}: 

\begin{quotation}
\noindent [...] a physically realistic spacetime {\em must} contain such singularities. [...] there exist causal, inextendible geodesics which are incomplete. [...] If a geodesic cannot be extended to a complete one (i.e. if its future endless continuation or its past endless continuation is of finite length), then either the particle suddenly ceases to exist or the particle suddenly springs into existence. In either case this can only happen if spacetime admits a ``singularity'' at the end (or the beginning) of the history of the particle.

\begin{flushright}
Kriele (1999), p. 383.
\end{flushright}
 
\end{quotation}
\vspace{0.5cm}

This statement and many similar ones found in the literature commit the elementary fallacy of confusing a model with the object being modeled. Space-time does not contain singularities. Some of our space-time models are singular. It is this incomplete character of the theory that prompt us to go beyond General Relativity in order to get a more comprehensive view of the gravitational phenomena. As it was very clear to Einstein, his general theory breaks down when the gravitational field of quantum objects starts to affect space-time. 

If General Relativity is not about singularities, what is it about? What is the ontology assumed by the theory? The answer is given by any of the several axiomatizations of GR that include explicit semantic axioms (Bunge 1967, Covarrubias 1993; see also the axioms presented above). The theory of General Relativity is about classical gravitational fields and the motion of material particles in them. No more, and no less, than what Einstein expressed in his letter to Sommerfeld of 1915 quoted in the second section of this paper.   

\section{Conclusion: no place for singularities}

I have argued in this paper that there are no physical space-time singularities. There cannot be, neither. Singularities are not physical entities, but limits of our ways of representing the world. There is no shame in that. General Relativity is beautiful enough as to admit theorems that can determine the conditions in which the theory cannot make consistent predictions. Excluding singularities from our language we shall pave the way to face the real problems posed to us: What is there inside black holes? What happened when the universe was under quantum effects? How General Relativity should be modified to account for this strange world?  

\section*{Appendix: an informal proof of the singularity theorem}

A congruence is a family of curves such that exactly one, and only one, time-like geodesic trajectory passes through each point $p\in E$. If the curves are smooth, a congruence defines a smooth time-like vector field on the space-time model. If $V^{a}$ is the time-like tangent vector to the congruence, we can write the {\em spatial part} of the metric tensor as:
\begin{equation}
	h_{ab}=g_{ab}+V_{a}V_{b}.
\end{equation}
 
For a given congruence of time-like geodesic we can define the {\em expansion}, {\em shear}, and {\em torsion} tensors as:

\begin{eqnarray}
\theta_{ab}&=&V_{(i;l)}h^{i}_{a}h^{l}_{b}, \\	
\sigma_{ab}&=&\theta_{ab}-\frac{1}{3}h_{ab}\theta,  \\
\omega_{ab}&=&h^{i}_{a}h^{l}_{b}V_{[i;l]}.  
\end{eqnarray}
Here, the {\em volume expansion} $\theta$ is defined as:

\begin{equation}
	\theta=\theta_{ab}h^{ab}=\nabla_{a}V^{a}=V^{a}_{\;\;\; ;a}.
\end{equation}

The rate of change of the volume expansion as the time-like geodesic curves in the congruence are moved along is given by the Raychaudhuri equation (Raychaudhuri 1955):

	\[\frac{d\theta}{d\tau}=-R_{ab}V^{a}V^{b}-\frac{1}{3}\theta^{2}-\sigma_{ab}\sigma^{ab}+\omega_{ab}\omega^{ab},
\]

or 

\begin{equation}
\frac{d\theta}{d\tau}=-R_{ab}V^{a}V^{b}-\frac{1}{3}\theta^{2}-2\sigma^{2}+2\omega^{2}.	
\end{equation}

We can use now the Einstein field equations to relate the congruence with the space-time curvature:

\begin{equation}
	R_{ab}V^{a}V^{b}= \kappa \left[T_{ab}V^{a}V^{b}+\frac{1}{2} T\right]. \label{cond-R}
\end{equation}

The term $T_{ab}V^{a}V^{b}$ represents the energy density measured by a tie-like observer with unit tangent four-velocity $V^{a}$. The weak energy condition then states that:
\begin{equation}
	T_{ab}V^{a}V^{b}\geq 0. \;\;\;\;{\rm WEC}
\end{equation}
 
A stronger condition is:
\begin{equation}
	T_{ab}V^{a}V^{b}+\frac{1}{2} T\geq 0. \;\;\;\;{\rm SEC}
\end{equation}
Notice that this condition implies, according to Eq. (\ref{cond-R}),
\begin{equation}
	R_{ab}V^{a}V^{b}\geq 0.
\end{equation}

We see, then, that the conditions of the Hawking-Penrose theorem imply that the focusing of the congruence yields:

\begin{equation}
	\frac{d\theta}{d\tau}\leq -\frac{\theta^{2}}{3}, \label{teor-sing}
\end{equation}
where we have used that both the shear and the rotation vanishes. Equation (\ref{teor-sing}) indicates that the volume expansion of the congruence must be necessarily decreasing along the time-like geodesic. Integrating, we get:
\begin{equation}
	\frac{1}{\theta}\geq \frac{1}{\theta_{0}}+ \frac{\tau}{3},
\end{equation}
where $\theta_{0}$ is the initial value of the expansion. Then, $\theta\rightarrow -\infty$ in a finite proper time $\tau\leq 3/\left|\theta_{0}\right|$. This means that once a convergence occurs in a congruence of time-like geodesics, a caustic must develop in the space-time model. The non space-like geodesics are in such a case inextendible and the space-time model singular.

\begin{acknowledgements}
An earlier version of this paper was presented in the Symposium Mario Novello's 70th Anniversary, held in Rio de Janeiro in 2012. I thank Mario Bunge, Mario Novello, Santiago E. Perez-Bergliaffa, and Daniela P\'erez for stimulating discussions. My work is partially supported by grant PIP 0078 (CONCET).
\end{acknowledgements}


\begin{thebibliography}{99}

\bibitem{Bunge0}
M. Bunge (1967).
\newblock {\em Foundations of Physics}.
\newblock Berlin-Heidelberg-New York: Springer-Verlag.

\bibitem{Bunge1}
M. Bunge (1974a). {\em Treatise of Basic Philosophy. Semantics I: Sense and Reference}. Dordrecht: Reidel.

\bibitem{Bunge2}
M. Bunge (1974b).
\newblock {\em Treatise of Basic Philosophy. Semantics II: Interpretation and Truth}.
\newblock Dordrecht: Reidel

\bibitem{Bunge1977}
M. Bunge (1977). {\itshape Ontology I: The Furniture of the World}. Dordrecht: Kluwer.

\bibitem{Bunge3}
M. Bunge (2006).
\newblock {\em Chasing Reality. Strife over Realism}.
\newblock Toronto: University of Toronto Press.


\bibitem{Clarke1993}
C.J.S. Clarke (1993).
\newblock {\em Analysis of Space-Time Singularities}.
\newblock Cambridge: Cambridge University Press.

\bibitem{Covarrubias}
G.M. Covarrubias (1993).
\newblock {An axiomatization of General Relativity}.
\newblock {\em International Journal of Theoretical Physics}, {\bf 32}, 2135-2154.

\bibitem{Earman1995}
J. Earman (1995).
\newblock {\em Bangs, Crunches, Whimpers, and Shrieks. Singularities and Acausalities in Relativistic Spacetimes}.
\newblock Oxford: Oxford University Press.

\bibitem{Einstein1916}
A. Einstein (1916).
\newblock N\"aherungweise Integration der Feldgleichungen der Gravitation. {\em Preussische Akademie der Wissenschaften}, 688-696.

\bibitem{E-R}
A. Einstein and N. Rosen (1935).
\newblock {The particle problem in the General Theory of Relativity}.
\newblock {\em Physical Review}, {\bf 48}, 73-77.


\bibitem{Godel}
K. G\"odel (1931).
\newblock {\"Uber formal unentscheidbare S\"atze der Principia Mathematica und verwandter Systeme}.
\newblock {\em Monatshefte f\"ur Mathematik und Physik}, {\bf 38}, 173-198.


\bibitem{H-P}
S.W. Hawking and R. Penrose (1970).
\newblock {The singularities of gravitational collapse and cosmology}.
\newblock {\em Proc. R. Soc. London A}, {\bf 314}, 529-548.

\bibitem{Hawking-73-Cambridge}
S.W. Hawking and G.F.R. Ellis (1973).
\newblock {\em The Large-Scale Structure of Space-Time}.
\newblock Cambridge: Cambridge University Press.

\bibitem{Kriele}
M. Kriele (1999).
\newblock {\em Spacetime: Foundations of General Relativity and Differential Geometry }.
\newblock Berlin-Heidelberg-New York: Springer.

\bibitem{Martin}
R.M. Martin (1978).
\newblock {\em Semiotics and Linguistic Structure}.
\newblock Albany: State University of New York Press.

\bibitem{Mink}
H. Minkowski (1908).
\newblock Lecture ``Raum und Zeit, 80th Versammlung Deutscher Naturforscher (K$\ddot{o}$ln, 1908)'', {\em Physikalische Zeitschrift}, {\bf 10}, 75-88 (1909).

\bibitem{Gravitation}
C. Misner, K.S. Thorne, and J.A. Wheeler (1973).
\newblock {\em Gravitation}.
\newblock San Francisco: W.H. Freeman.


\bibitem{Bergliaffa1}
S.E. Perez-Bergliaffa, G.E Romero, and H. Vucetich (1993).
\newblock {Axiomatic foundations of nonrelativistic quantum mechanics: a realistic approach}.
\newblock {\em International Journal of Theoretical Physics}, {\bf 32}, 1507-1525.

\bibitem{Bergliaffa2}
S.E. Perez-Bergliaffa, G.E Romero, and H. Vucetich (1996).
\newblock {Axiomatic foundations of quantum mechanics revisited: the case for systems}.
\newblock {\em International Journal of Theoretical Physics}, {\bf 35}, 1805-1819.

\bibitem{Bergliaffa3}
S.E. Perez-Bergliaffa, G.E Romero, and H. Vucetich (1998).
\newblock {Toward an axiomatic pregeometry of space-time}.
\newblock {\em International Journal of Theoretical Physics}, {\bf 37}, 2281-2298.

\bibitem{Ray}
A.K. Raychaudhuri (1955).
\newblock {Relativistic cosmology}.
\newblock {\em Phys. Rev.}, {\bf 98}, 1123-1126.

\bibitem{Romero1}
G.E. Romero (2012a).
\newblock {Parmenides reloaded}.
\newblock {\em Foundations of Science}, {\bf 17}, 291-299.


\bibitem{Romero2}
G.E. Romero (2012b).
\newblock {From change to space-time: an Eleatic journey}.
\newblock {\em Foundations of Science}, DOI 10.1007/s10699-011-9272-5, in press.

\bibitem{Rovelli}
C. Rovelli (2004). \emph{Quantum Gravity}, Cambridge: Cambridge University Press.

\bibitem{Tarski}
A. Tarski (1983).
\newblock {\em Logic, Semantics, Metamathematics}.
\newblock Indianapolis: Hackett Pub. Co. Inc.

\bibitem{Dongen}
J. van Dongen (2010).
\newblock {\em Einstein's Unification}.
\newblock Cambridge: Cambridge University Press. 

\bibitem{Vizgin}
V.P. Vizgin (1994).
\newblock {\em Unified Field Theories in the First Third of the 20th Century}.
\newblock Basel: Birkh\"auser Verlag.














































\end{thebibliography}



\newpage

\section*{Gustavo E. Romero} Full Professor of Relativistic Astrophysics at the University of La Plata and Chief Researcher of the National Research Council of Argentina. A former President of the Argentine Astronomical Society, he has published more than 250 papers on astrophysics, gravitation, and the foundation of physics, and 8 books. His main current interest is on black hole physics and ontological problems of space-time theories.

\end{document}